\documentclass[conference]{IEEEtran}
\IEEEoverridecommandlockouts
\usepackage{amsmath,graphicx}
\usepackage{epstopdf, latexsym, amssymb, cite, ushort, mathtools, caption, url, pgfplots, tikzscale}
\usepackage{booktabs}
\usepackage{ushort}
\usepackage{calrsfs}
\DeclareMathAlphabet{\pazocal}{OMS}{zplm}{m}{n}
\usepackage{makecell}
\usepackage{array}
\usepackage{textcomp}
\usepackage{calc}
\usepackage{tikz}
\usetikzlibrary{shapes,arrows}
\usetikzlibrary{calc}
\usetikzlibrary{positioning}
\usetikzlibrary{arrows}
\usetikzlibrary{arrows.meta}
\usetikzlibrary{arrows,scopes}
\usetikzlibrary{backgrounds}
\usetikzlibrary{patterns}
\usepackage{xfrac} 
\usepackage{nicefrac}  
\usepackage{upgreek} 
\usepackage{cases} 
\usepackage{subcaption} 
\usepackage{siunitx}
\usepackage{empheq}
\usepackage{xcolor}
\def\BibTeX{{\rm B\kern-.05em{\sc i\kern-.025em b}\kern-.08em
    T\kern-.1667em\lower.7ex\hbox{E}\kern-.125emX}}

\DeclareMathOperator {\expon}{exp}
\DeclareMathOperator {\diag}{diag}
\DeclareMathOperator {\Diag}{Diag}

\DeclareMathOperator {\Exp}{E}

\graphicspath{{../../../figures/matlab2tikz/}}

\allowdisplaybreaks

\def\a{\mathbf{a}}

\def\x{\mathbf{x}}
\def\y{\mathbf{y}}
\def\v{\mathbf{v}}
\def\A{\mathbf{A}}

\def\w{\mathbf{w}}
\def\h{\mathbf{h}}

\def\p{{\mathbf{p}}}
\def\0{{\mathbf{0}}}

\def\A{\mathbf{A}}

\def\P{\mathbf{P}}

\def\bPsi{\boldsymbol{\Psi}}

\def\I{\mathbf{I}}

\def\bGamma{\mathbf{\Gamma}}

\renewcommand{\-}{\hskip 0.13em{-}\hskip 0.13em}

\newcommand*{\transp}{{\scriptscriptstyle{T}}}
\newcommand*{\herm}{{\scriptscriptstyle{H}}}
\newcommand*{\inv}{{\scriptscriptstyle{-1}}}

\tikzset{%
  block/.style    = {draw, thick, rectangle, minimum height = 2.5em,
    minimum width = 2.5em},
  sum/.style      = {draw, circle, node distance = 3em}, 
  input/.style    = {coordinate}, 
  output/.style   = {coordinate} 
}
\tikzset{
  -|-/.style={
    to path={
      (\tikztostart) -| ($(\tikztostart)!#1!(\tikztotarget)$) |- (\tikztotarget)
      \tikztonodes
    }
  },
  -|-/.default=0.5,
  |-|/.style={
    to path={
      (\tikztostart) |- ($(\tikztostart)!#1!(\tikztotarget)$) -| (\tikztotarget)
      \tikztonodes
    }
  },
  |-|/.default=0.5,
}
\tikzset{microphone/.style={black,circle,draw,fill=white,scale=1}}
\tikzset{conjunction/.style={black,circle,draw,fill=black,scale=0.25}}

\definecolor{darkGray}{rgb}{0.7,0.7,0.7}
\definecolor{lightGray}{rgb}{0.8,0.8,0.8}
\definecolor{lightlightGray}{rgb}{0.95,0.95,0.95}

\pgfdeclarelayer{bg}    
\pgfsetlayers{bg,main}  

\definecolor{mycolor1}{rgb}{0.00000,0.70000,1.00000}
\definecolor{mycolor2}{rgb}{0.00000,0.70000,1.00000}
\newlength\fheight 
\newlength\fwidth

\newcolumntype{M}[1]{>{\centering\arraybackslash}m{#1}}
\newcolumntype{N}{@{}m{0pt}@{}}

\begin{document}

\title{Instantaneous PSD Estimation \\ for Speech Enhancement based on \\ Generalized Principal Components
\thanks{This work was carried out at the ESAT Laboratory of KU Leuven, in the frame of KU Leuven internal fund C2-16-00449; VLAIO O\&O Project no. HBC.2017.0358; EU FP7-PEOPLE Marie Curie Initial Training Network funded by the European Commission under Grant Agreement no. 316969; the European Union's Horizon 2020 research and innovation program/ERC Consolidator Grant no. 773268. This paper reflects only the authors' views and the Union is not liable for any use that may be made of the contained information.}
}

\author{\IEEEauthorblockN{Thomas Dietzen, Marc Moonen, Toon van Waterschoot}
\IEEEauthorblockA{\textit{Dept. of Electrical Engineering (ESAT)}\\ 
\textit{STADIUS Center for Dynamical Systems, Signal Processing and Data Analytics}\\
\textit{KU Leuven}\\
Leuven, Belgium \\
\{thomas.dietzen, marc.moonen, toon.vanwaterschoot\}@esat.kuleuven.be}
}

\maketitle

\begin{abstract}
Power spectral density (PSD) estimates of various microphone signal components are essential to many speech enhancement procedures.
As speech is highly non-nonstationary, performance improvements may be gained by maintaining time-variations in PSD estimates.
In this paper, we propose an instantaneous PSD estimation approach based on generalized principal components. 
Similarly to other eigenspace-based PSD estimation approaches, we rely on recursive averaging in order to obtain a microphone signal correlation matrix estimate to be decomposed. 
However, instead of estimating the PSDs directly from the temporally smooth generalized eigenvalues of this matrix, yielding temporally smooth PSD estimates, we propose to estimate the PSDs from newly defined instantaneous generalized eigenvalues, yielding instantaneous PSD estimates.
The instantaneous generalized eigenvalues are defined from the generalized principal components, i.e. a generalized eigenvector-based transform of the microphone signals.
We further show that the smooth generalized eigenvalues can be understood as a recursive average of the instantaneous generalized eigenvalues.
Simulation results comparing the multi-channel Wiener filter (MWF) with smooth and instantaneous PSD estimates indicate better speech enhancement performance for the latter.
A MATLAB implementation is available online.
\end{abstract}


\section{Introduction}

In speech enhancement \cite{loizou2007speech, doclo2010acoustic, gannot2017consolidated},  recorded microphone signals constitute a mixture of speech, reverberation and noise.
In order to enhance the mixture, many approaches rely on power spectral density (PSD)
estimates of the various mixture components.

While the problem of PSD estimation has attracted much interest  \cite{loizou2007speech, gannot2017consolidated, Cohen2003, Hendriks2008, Kamkar2011, KuklasinskiDJJ16, SchwartzGH16b, Braun18TASLP, KodrasiD18, Koutrouvelis18, dietzen19TASLP_SQRTpub} in speech enhancement, somewhat less attention \cite{Cohen2003, Hendriks2008, Kamkar2011, dietzen19TASLP_SQRTpub} is paid to the temporal behavior of PSD estimates. 
As the PSD is a statistical property defined by means of an expectation operator, its estimation typically involves temporal averaging, which approximates the expectation and requires tuning. 
Note that while temporal averaging lacks practical alternatives, it causes temporal smoothing and hence may be considered non-ideal in case of speech signals, which are highly non-stationary.
Indeed, non-stationarity of speech may even be explicitly exploited in a number of speech enhancement approaches \cite{loizou2007speech, Li2008, Wang2009, Narayanan2013, dietzen19TASLP_ISCLPpub}, such that quickly time-varying PSD estimates potentially yield a better performance than slowly time-varying PSD estimates. 
In literature, quickly time-varying PSD estimates are commonly based on short-term statistics, e.g., the local minima of the smoothed  microphone signal spectrum \cite{Cohen2003} or short-term temporal correlations \cite{Hendriks2008, Kamkar2011}.
In \cite{dietzen19TASLP_SQRTpub}, we have proposed to restore non-stationarities by desmoothing the generalized eigenvalues of the temporally smooth microphone signal correlation matrix estimate.

In this paper,
 we propose a multi-microphone eigenspace-based \textit{instantaneous} PSD\footnote{Strictly speaking, the term 'PSD' may be said to be inadequate for the instantaneous quantities estimated in this paper, as our approach partly bypasses the use of an expectation or its approximation by means of temporal averaging. Nonetheless, due to the strong relation to expectation-based PSD estimation, we prefer to maintain the terminology.} estimation approach based on generalized principal components. 
Similarly to other eigenspace-based PSD estimation approaches \cite{KuklasinskiDJJ16, Hendriks2008,  KodrasiD18, dietzen19TASLP_SQRTpub}, we rely on recursive averaging in order to obtain a microphone signal correlation matrix estimate to be decomposed. 
However, instead of estimating the PSDs directly from the temporally \textit{smooth} generalized eigenvalues of this matrix, yielding temporally smooth PSD estimates, we propose to estimate the PSDs from newly defined \textit{instantaneous} generalized eigenvalues, yielding instantaneous PSD estimates.
Here, the instantaneous generalized eigenvalues are defined from the generalized principal components, i.e. a generalized eigenvector-based transform of the microphone signals.
As to be shown, the smooth generalized eigenvalues can be understood as a recursive average of the  newly defined instantaneous generalized eigenvalues.
Simulation results comparing the speech enhancement performance of the multi-channel Wiener filter (MWF) with smooth and instantaneous PSD estimates indicate better performance for the latter. 
A MATLAB implementation and audio examples are available online \cite{eusipco2020Code}. 

In Sec. \ref{sec:sm}, we present the signal model.
In Sec. \ref{sec:MWF}, we briefly review the MWF, which serves as an example for the application of PSD estimates and is used to evaluate PSD estimates in this paper.
Eigenspace-based PSD estimation is discussed in Sec. \ref{sec:GEVD}, where we outline an implementation yielding smooth PSD estimates and propose the alternative approach yielding instantaneous PSD estimates.
Both implementations are evaluated in Sec. \ref{sec:sim}.

\section{Signal Model}
\label{sec:sm}

We employ the following notation: vectors are denoted by lower-case boldface letters, matrices by upper-case boldface letters, $\mathbf{I}$ denotes the identity matrix, $\mathbf{A}^\transp$, $\mathbf{A}^\herm$, $\Exp[\mathbf{A}]$, and $\Vert\A\Vert_\textsl{F}$ denote the transpose, the complex conjugate transpose, the expected value, and the Frobenius norm of the matrix $\A$.
The operation  $\diag[\A]$ creates a column vector from the diagonal elements of the matrix $\A$, while $\Diag[\a]$ creates a diagonal matrix from the elements of the vector $\a$.
The exponential function with argument $a$ is denoted by $\expon[a]$.

In the short-time Fourier transform (STFT) domain, with $m$, $l$, and $k$ indexing the microphone, the frame, and the frequency bin, respectively, and $M$ the number of microphones, let the microphone signals be denoted by $y_m(l,k) \in \mathbb{C}$ with $m=1,\dots,M$.
As we treat all frequency bins independently, the frequency bin index is omitted in the following.
We define the stacked microphone signal vector $\y(l) \in \mathbb{C}^{M}$,
\begin{align}
\y(l) &= 
\begin{pmatrix}
y_1(l) & \cdots & y_{M}(l)
\end{pmatrix}^{\transp}, 
\label{eq:sm:y_stacked}
\end{align}
composed of the reverberant speech component $\x(l)$ originating from a single point source and the noise component $\v(l)$, 
\begin{align}
{\y}(l) &=\x(l) + \v(l). 
\label{eq:sm:y_decomp}
\end{align}
The reverberant speech component $\x(l)$ may be decomposed into the early component ${\x}_{\textsl{e}}(l)$ containing the direct component and early reflections, and the late reverberant component ${\x}_{\ell}(l)$ containing late reflections, i.e. 
\begin{align}
{\x}(l) &= {\x}_{\textsl{e}}(l)+ {\x}_{\ell}(l),
\label{eq:sm:xn_decomp}
\end{align}
which are assumed to have distinct spatial properties as outlined below.
Early reflections are assumed to arrive within the same frame,
with the early components in ${\x}_{\textsl{e}}(l)$ related by the relative early transfer functions (RETFs) in $\h \in \mathbb{C}^{M}$, i.e.
\begin{align}
{\x}_{\textsl{e}}(l) &=  \h s(l), 
\label{eq:sm:xne_RETF}
\end{align}
Here, $\h$ is assumed to be relative to the first microphone, i.e. $h_1 = 1$, and $s(l) = {\x}_{\textsl{e}|1}(l)$ denotes the early component in the first microphone, in the following referred to as early speech source image. 
We consider $\h$ to be known or previously estimated \cite{gannot2017consolidated, dietzen19TASLP_SQRTpub, MarkovichGolanShmulik2015Paot}.
We assume that ${\x}_{\textsl{e}}(l)$, ${\x}_{\ell}(l)$, and ${\v}(l)$ are mutually uncorrelated  \cite{KuklasinskiDJJ16, SchwartzGH16b, Braun18TASLP, KodrasiD18, Koutrouvelis18, dietzen19TASLP_SQRTpub}.
Let $\bPsi_{y}(l) = \Exp[\y(l)\y^\herm(l)] \in \mathbb{C}^{M \times M}$ denote the microphone signal correlation matrix, and let ${\bPsi}_{x_\textsl{e}}(l)$, ${\bPsi}_{x_\ell}(l)$, and ${\bPsi}_{v}(l)$ be similarly defined.
With (\ref{eq:sm:y_decomp})--(\ref{eq:sm:xne_RETF}), we then find
\begin{align}
{\bPsi}_{y}(l)  = {\bPsi}_{x_\textsl{e}}(l) + {\bPsi}_{x_\ell}(l) + {\bPsi}_{v}(l),
\label{eq:sm:Psiy}
\end{align}
wherein ${\bPsi}_{x_\textsl{e}}(l)$ has rank one and is expressed by 
\begin{align}
{\bPsi}_{x_\textsl{e}}(l)  &=  \varphi_{s}(l){\h}{\h}^\herm, \label{eq:sm:rankOneRETF}
  \end{align}
 with $\varphi_{s}(l)$ denoting the PSD of the early speech source image $s(l)$.
 Assuming that ${\x}_{\ell}(l)$ and $\v(l)$ may be modeled as diffuse \cite{KuklasinskiDJJ16, SchwartzGH16b, Braun18TASLP, KodrasiD18,  Koutrouvelis18, jacobsen2000coherence} with coherence matrix $\mathbf{\Gamma}  \in \mathbb{C}^{M \times M}$, which may be computed from the microphone array geometry \cite{jacobsen2000coherence} and is therefore considered to be known, we may write ${\bPsi}_{x_\ell}(l) + {\bPsi}_{v}(l)$ as
 \begin{align}
 {\bPsi}_{x_\ell}(l) + {\bPsi}_{v}(l) &= {\varphi}_{\textsl{d}}(l)\mathbf{\Gamma},  \label{eq:sm:Psixl}\\
  \text{with}\quad{\varphi}_{\textsl{d}}(l) &= \varphi_{x_{\ell}}(l) + \varphi_{v}(l),
  \label{eq:sm:phid}
 \end{align}
 and $\varphi_{x_{\ell}}(l)$ and $\varphi_{v}(l)$ denoting the PSD of the  late reverberant component ${\x}_{\ell}(l)$ and the noise component $\v(l)$, respectively.
With $s(l)$ representing speech, and in particular if $\v(l)$ represents babble noise,
both PSDs  ${\varphi}_{s}(l)$ and ${\varphi}_{\textsl{d}}(l)$ may be considered highly non-stationary, while the associated coherence matrices $\h\h^\herm$ and  $\mathbf{\Gamma}$ are often considered time-invariant  \cite{KuklasinskiDJJ16, SchwartzGH16b, Braun18TASLP, KodrasiD18}.

In the remainder, as we mostly consider the single frame $l$ only, we also drop the frame index for conciseness and refer back to it only where necessary, namely when we differentiate the frames $l$ and $l-1$ in recursive equations.

\section{Multi-Channel Wiener Filter}
\label{sec:MWF}

PSD estimates are used in a variety of speech enhancement procedures. 
In this paper, we evaluate our PSD estimation approach in Sec. \ref{sec:sim} by means of the MWF, which is therefore briefly summarized below.

The MWF ${\w}_{\tiny\text{MWF}}$ is obtained \cite{doclo2010acoustic, gannot2017consolidated} by minimizing the expected error between the filter output and the early speech source image, i.e. 
\begin{align}
{\w}_{\tiny\text{MWF}} &= \underset{\w}{\arg\min}\,
	\Exp\bigl[|\w^\herm\y - s|^2\bigr]\nonumber\\ 
	&= \varphi_{s}\bPsi_{y}^\inv \h.
\label{eq:...}
\end{align}
It is well known that the MWF can be decomposed \cite{doclo2010acoustic, gannot2017consolidated} into a minimum variance distortionless response (MVDR) beamformer and a spectral gain as
\begin{align}
{\w}_{\tiny\text{MWF}} = \underbrace{\dfrac{\bGamma^\inv\h}{\h^\herm\bGamma^\inv\h} \vphantom{ \sum_*^*} }_{\substack{ \text{MVDR} \\  \text{beamformer}}} 
\cdot 
\underbrace{\dfrac{\varphi_{s}}{\varphi_{s} + \varphi_{\textsl{d}} \h^\herm\bGamma^\inv\h } \vphantom{ \sum_*^*} }_{ \text{ spectral gain}}.
\label{eq:MWFgain}
\end{align}
Hence, if both $\bGamma$ and $\h$ are assumed to be known or previously estimated, the problem of implementing the MWF reduces to estimating the PSDs $\varphi_{s}$ and $\varphi_{\textsl{d}}$.
If, on the one hand, the PSD estimates to be obtained are slowly time-varying, the spectral gain will contribute to speech enhancement mostly through variations across frequency. 
If, on the other hand, instantaneous PSD estimates are obtained, the spectral gain will vary across both frequency and time and thereby act as a spectro-temporal mask \cite{Narayanan2013, Li2008, Wang2009}.

\section{Eigenspace-based PSD Estimation}
\label{sec:GEVD}

Multi-microphone PSD estimation is commonly based on the spatial properties defined in (\ref{eq:sm:xne_RETF})--(\ref{eq:sm:phid}), which may be exploited in an eigenspace decomposition \cite{KuklasinskiDJJ16,  KodrasiD18, dietzen19TASLP_SQRTpub}.
In Sec. \ref{sec:eigenspaceModel}, we first introduce an eigenspace model of ${\bPsi}_{y}$ and $\mathbf{\Gamma}$. In Sec. \ref{sec:PSDestimation}, we outline how PSD estimates may be obtained given an eigenvalue and an eigenspace basis estimate. 
In Sec. \ref{sec:smoothEigenvalues}, we consider an implementation based on temporally smooth eigenvalues, and in Sec. \ref{sec:instantPC}, we propose an implementation based on instantaneous generalized principal components. 

\subsection{Eigenspace Model}
\label{sec:eigenspaceModel}

We define the generalized eigenvalue decomposition (GEVD) \cite{MarkovichGolanShmulik2015Paot, KuklasinskiDJJ16,  KodrasiD18, dietzen19TASLP_SQRTpub} of ${\bPsi}_{y}$ and the diffuse coherence matrix $\mathbf{\Gamma}$, cf. (\ref{eq:sm:Psixl}), i.e.
\begin{align}
{\bPsi}_{y}\P &= \mathbf{\Gamma}\P\Diag[{\boldsymbol{\lambda}}_{y}],
\label{eq:smoothGEVD:PsiyHatGEVD}
\end{align}
where ${\boldsymbol{\lambda}}_{y} \in \mathbb{R}^{M}$ comprises the generalized eigenvalues $\lambda_{y|m}$, and the columns $\p_m$ of $\P \in \mathbb{C}^{M\times M}$ comprise the associated generalized eigenvectors.
The generalized eigenvectors in $\P$ are uniquely defined up to a scaling factor and, for any factorization $\mathbf{\Gamma} = \mathbf{\Gamma}^{\nicefrac{1}{2}}\mathbf{\Gamma}^{\nicefrac{\herm}{2}}$, may be chosen such that $\mathbf{\Gamma}^{\nicefrac{\herm}{2}}\P$ becomes unitary due to  ${\bPsi}_{y}$ and $\mathbf{\Gamma}$ being Hermitian.
The matrices ${\bPsi}_{y}$ and $\mathbf{\Gamma}$  are then diagonalized by
\begin{align}
 \P^\herm{\bPsi}_{y}\P &= \Diag[{\boldsymbol{\lambda}}_{y}],  \label{eq:smoothGEVD:eigvalues}\\
\P^\herm\mathbf{\Gamma}\P &= \I,
\label{eq:smoothGEVD:P_scaling}
\end{align}
cf. also (\ref{eq:smoothGEVD:PsiyHatGEVD}).

While the eigenspace basis $\P$ varies with the spatial coherence matrices $\h\h^\herm$ and $\mathbf{\Gamma}$ only and is therefore time-invariant in the assumed spatially stationary scenario, the generalized eigenvalues in ${\boldsymbol{\lambda}}_{y}$ vary with the PSDs $\varphi_s$ and $\varphi_\textsl{d}$ and hence over time.
Using (\ref{eq:sm:Psiy}) and (\ref{eq:sm:Psixl}) in (\ref{eq:smoothGEVD:eigvalues})--(\ref{eq:smoothGEVD:P_scaling}) yields
\begin{align}
\Diag[{\boldsymbol{\lambda}}_y] &= \Diag[{\boldsymbol{\lambda}}_{x_\textsl{e}}] +   \Diag[{\boldsymbol{\lambda}}_{\textsl{d}}], \label{eq:decompGEVD:LambdayHat}\\
\text{with}\quad \Diag[{\boldsymbol{\lambda}}_{x_\textsl{e}}] &= \P^\herm{\bPsi}_{x_\textsl{e}}\P, \label{eq:decompGEVD:Lambdaxe}\\
 \Diag[{\boldsymbol{\lambda}}_{\textsl{d}}] &= {\varphi}_{\textsl{d}}\I.
\end{align}
In (\ref{eq:decompGEVD:Lambdaxe}), ${\bPsi}_{x_\textsl{e}}$ and therefore $\Diag[{\boldsymbol{\lambda}}_{x_\textsl{e}}]$ have rank one. 
Provided that the generalized eigenvalues and eigenvectors are sorted such that  ${{\lambda}}_{y|1}$ is the largest generalized eigenvalue, ${\boldsymbol{\lambda}}_{x_\textsl{e}}$ hence takes the form 
\begin{align}
{\boldsymbol{\lambda}}_{x_\textsl{e}} &=
\begin{pmatrix} 
{{\lambda}}_{{x_\textsl{e}|1}} & 0 & \cdots & 0
\end{pmatrix}^\transp. \label{eq:decompGEVD:PPsiP}
\end{align}
From (\ref{eq:decompGEVD:LambdayHat})--(\ref{eq:decompGEVD:PPsiP}) it then follows that ${{\lambda}}_{y|1} = {{\lambda}}_{x_\textsl{e}|1} + {\varphi}_{\textsl{d}}$ and ${{\lambda}}_{y|m} = {\varphi}_{\textsl{d}}$ for $m > 1$ \cite{KodrasiD18}.

\subsection{Eigenspace-based PSD Estimation}
\label{sec:PSDestimation}

Assume that an estimate $\hat{\bPsi}_{y}$ is available, from which the eigenvalue and eigenspace basis estimates $\hat{\boldsymbol{\lambda}}_{y}$ and $\hat{\P}$ are obtained.
Further, assume that the RETF ${\h}$ is known or previously estimated.
Estimates of  $\hat{\varphi}_{s}$ and $\hat{\varphi}_{\textsl{d}}$ can then be obtained in the following manner.

Given $\hat{\boldsymbol{\lambda}}_{y}$, we first obtain $\hat{\varphi}_{\textsl{d}}$ and $\hat{\lambda}_{x_\textsl{e}|1}$ according to   (\ref{eq:decompGEVD:LambdayHat})--(\ref{eq:decompGEVD:PPsiP}) \cite{KodrasiD18} as
\begin{align}
\hat{\varphi}_{\textsl{d}} &=
\dfrac{1}{M-1}{\sum_{m=2}^M\hat{{\lambda}}_{y|m}}, \label{eq:phid_hat}\\
\hat{\lambda}_{x_\textsl{e}|1} &=
\hat{{\lambda}}_{y|1} - \hat{\varphi}_{\textsl{d}}. \label{eq:lambda_xe1}
\end{align}
where the averaging in (\ref{eq:phid_hat}) accounts for modeling and estimation errors and (\ref{eq:lambda_xe1}) is guaranteed non-negative.
Noting that $\hat{\P}^\inv =  \hat{\P}^\herm\mathbf{\Gamma}$ according to (\ref{eq:smoothGEVD:P_scaling}), we can define a rank-one estimate $\hat{\bPsi}_{x_{\textsl{e}}}$ \cite{dietzen19TASLP_SQRTpub} as 
\begin{align}
\hat{\bPsi}_{x_\textsl{e}} &=  \mathbf{\Gamma}\hat{\P}
\Diag[\hat{\boldsymbol{\lambda}}_{x_\textsl{e}}] 
\hat{\P}^\herm\mathbf{\Gamma} \nonumber\\
&=  \hat{{\lambda}}_{x_\textsl{e}|1}\mathbf{\Gamma}\hat{\p}_{1} \hat{\p}_{1}^\herm\mathbf{\Gamma}
\end{align}
with $\hat{\boldsymbol{\lambda}}_{x_\textsl{e}}$ similar to (\ref{eq:decompGEVD:PPsiP}). 
An estimate $\hat{\varphi}_{s}$ may then be obtained by minimizing the difference\footnote{Since $h_1 = 1$, cf. Sec. \ref{sec:sm}, one may alternatively obtain an estimate $\hat{\varphi}_{s}$ directly from the upper left element of $\hat{\bPsi}_{x_\textsl{e}}$ \cite{dietzen19TASLP_SQRTpub}.
During speech pauses, however, where $\hat{\bPsi}_{x_\textsl{e}}$ deviates from zero due to modeling and estimation errors only, the estimator in (\ref{eq:sourceimagePSDestimate}) is more robust.
} between $\hat{\bPsi}_{x_\textsl{e}}$ and $\varphi_s{\h}{\h}^\herm$ according to (\ref{eq:sm:rankOneRETF}) \cite{dietzen19TASLP_SQRTpub}, i.e.
\begin{align}
\hat{\varphi}_{s} &= \underset{\varphi_{s}}{\arg\min}\,
	\lVert  \varphi_s{\h}{\h}^\herm -  \hat{\bPsi}_{x_\textsl{e}}   \rVert ^2_\textsl{F}      \nonumber\\ 
	&=  \hat{{\lambda}}_{x_\textsl{e}|1} \left\lvert \dfrac{{\h}^\herm \mathbf{\Gamma}\hat{\p}_{1}}{ {\h}^\herm{\h}} \right\rvert^2.
\label{eq:sourceimagePSDestimate}
\end{align}
Note that the temporal characteristics of the estimates $\hat{\varphi}_{s}$ and $\hat{\varphi}_{\textsl{d}}$ directly depend upon the temporal characteristics of $\hat{\boldsymbol{\lambda}}_{y}$. 

\subsection{Smooth Eigenvalue-based Implementation}
\label{sec:smoothEigenvalues}
A temporally smooth estimate of ${\bPsi}_{y} = \Exp[\y\y^\herm]$, in the following denoted by $\hat{\bPsi}_{y|\textsl{sm}}$, is typically obtained by recursively averaging $\y\y^\herm$ using some pre-defined forgetting factor $\zeta \in (0,\,1)$, namely by
\begin{align}
\hat{\bPsi}_{y|\textsl{sm}}(l) = \zeta \hat{\bPsi}_{y| \textsl{sm}}(l\-1) + (1\-\zeta)\y(l)\y^\herm(l).\label{eq:smoothGEVD:PsiyHat}
\end{align}
The forgetting factor $\zeta$ may be expressed in terms of a time constant $\tau$ as 
\begin{align}
\zeta = \expon\bigl[{\nicefrac{\displaystyle -R}{ \displaystyle f_\textsl{s} \tau}}\bigr],\label{eq:zeta}
\end{align}
where $R$ is the STFT frame shift in samples, $f_\textsl{s}$ is the sampling rate and $\tau$ may be thought of as an equivalent window length.

Given $\hat{\bPsi}_{y|\textsl{sm}}$, we can perform the GEVD $\hat{\bPsi}_{y|\textsl{sm}}\hat{\P} = \mathbf{\Gamma}\hat{\P}\Diag[\hat{\boldsymbol{\lambda}}_{y|\textsl{sm}}]$ similar to (\ref{eq:smoothGEVD:PsiyHatGEVD})--(\ref{eq:smoothGEVD:P_scaling}) in each frame $l$.
Here, $\hat{\P}$ slightly fluctuates over time due to modeling and estimation errors (while ${\P}$ itself is time-invariant, cf. Sec. \ref{sec:eigenspaceModel}), and $\hat{\boldsymbol{\lambda}}_{y|\textsl{sm}}$ is a smooth estimate of  ${\boldsymbol{\lambda}}_{y}$.
Consequently, if we estimate the PSDs ${\varphi}_{s}$ and ${\varphi}_{\textsl{d}}$ directly from $\hat{\boldsymbol{\lambda}}_{y|\textsl{sm}}$ according to Sec. \ref{sec:PSDestimation}, we obtain  equally smooth estimates  $\hat{\varphi}_{s|\textsl{sm}}$ and $\hat{\varphi}_{\textsl{d}|\textsl{sm}}$.
Note that in order to span all $M$ eigenspace dimensions and hence to obtain a meaningful decomposition,  $\hat{\bPsi}_{y|\textsl{sm}}$ needs to be well-conditioned, and so  $\tau$ should scale with $M$ and must be sufficiently large.

\subsection{Instantaneous Principal Component-based Implementation}
\label{sec:instantPC}
In order to obtain instantaneous eigenspace-based PSD estimates while still relying on recursive averaging as in (\ref{eq:smoothGEVD:PsiyHat}) with a sufficiently large time constant $\tau$, we propose to compute instantaneous generalized eigenvalues $\hat{\boldsymbol{\lambda}}_{y | \textsl{inst}}$ based on generalized principal components instead of using the smooth generalized eigenvalues $\hat{\boldsymbol{\lambda}}_{y|\textsl{sm}}$ directly.

In order to introduce the generalized principal components and establish its relation to the generalized eigenvalues, let us reconsider the GEVD in (\ref{eq:smoothGEVD:PsiyHatGEVD})--(\ref{eq:smoothGEVD:P_scaling}).
From the generalized eigenvectors in $\P$, we can define the generalized principal components of $\y$ as
\begin{align}
\boldsymbol{\vartheta} = \P^\herm\y. \label{eq:PC}
\end{align}
Note that with $\bPsi_{y} = \Exp[\y\y^\herm]$, the generalized principal components in (\ref{eq:PC}) are related to the generalized eigenvalues in (\ref{eq:smoothGEVD:eigvalues}) by 
\begin{align}
{\boldsymbol{\lambda}}_{y} = \diag\bigl[\Exp[\boldsymbol{\vartheta}\boldsymbol{\vartheta}^\herm]\bigr].
\end{align}
Now, assume that we have obtained $\hat{\bPsi}_{y|\textsl{sm}}$ and its generalized eigenvectors in   $\hat{\P}$ as described in Sec. \ref{sec:smoothEigenvalues}.
Then, with $\hat{\boldsymbol{\vartheta}} = \hat{\P}\y$, we define the instantaneous generalized eigenvalues 
\begin{align}
\hat{\boldsymbol{\lambda}}_{y | \textsl{inst}} 
&= \diag[\hat{\boldsymbol{\vartheta}}\hat{\boldsymbol{\vartheta}}^\herm], \label{eq:eigenvaluesPC}
\end{align}
which maintain non-stationarities as they directly depend on the microphone signal $\y$, cf. (\ref{eq:PC}).
Based on $\hat{\boldsymbol{\lambda}}_{y | \textsl{inst}}$, we can then obtain instantaneous PSD estimates $\hat{\varphi}_{s|\textsl{inst}}$ and $\hat{\varphi}_{\textsl{d}|\textsl{inst}}$ according to Sec. \ref{sec:PSDestimation}. 

Note that we may also establish a relation between the instantaneous generalized eigenvalues in (\ref{eq:eigenvaluesPC}) and the smooth generalized eigenvalues obtained in Sec. \ref{sec:smoothEigenvalues}. 
With $\hat{\boldsymbol{\lambda}}_{y | \textsl{sm}} =  \diag[\hat{\P}^\herm\hat{\bPsi}_{y| \textsl{sm}}\hat{\P}]$ according to (\ref{eq:smoothGEVD:eigvalues}), inserting $\hat{\bPsi}_{y| \textsl{sm}}$ from (\ref{eq:smoothGEVD:PsiyHat}) and using (\ref{eq:PC}), (\ref{eq:eigenvaluesPC}), we find 
\begin{align}
\hat{\boldsymbol{\lambda}}_{y | \textsl{sm}}(l) &=  \zeta\diag[\hat{\P}(l)^\herm  \hat{\bPsi}_{y| \textsl{sm}}(l\-1)    \hat{\P}(l)] \nonumber\\
&\quad 
+ (1\-\zeta)\hat{\boldsymbol{\lambda}}_{y | \textsl{inst}}(l), 
\end{align}
where any time variations in $\hat{\P}(l)$ are due to modeling and estimation errors only, cf. Sec. \ref{sec:smoothEigenvalues}, such that $\diag[\hat{\P}(l)^\herm  \hat{\bPsi}_{y| \textsl{sm}}(l\-1)  \hat{\P}(l)] \approx \hat{\boldsymbol{\lambda}}_{y | \textsl{sm}}(l\-1)$.
The smooth generalized eigenvalues $\hat{\boldsymbol{\lambda}}_{y | \textsl{sm}}$ therefore nearly correspond to a recursive average of the instantaneous generalized eigenvalues $\hat{\boldsymbol{\lambda}}_{y | \textsl{inst}}$.

\section{Simulations}
\label{sec:sim}

 \setlength\fwidth{7.85cm}
 \begin{figure}
\centering
\hspace*{-0.26cm}
    \setlength\fheight{16cm} 
 \input{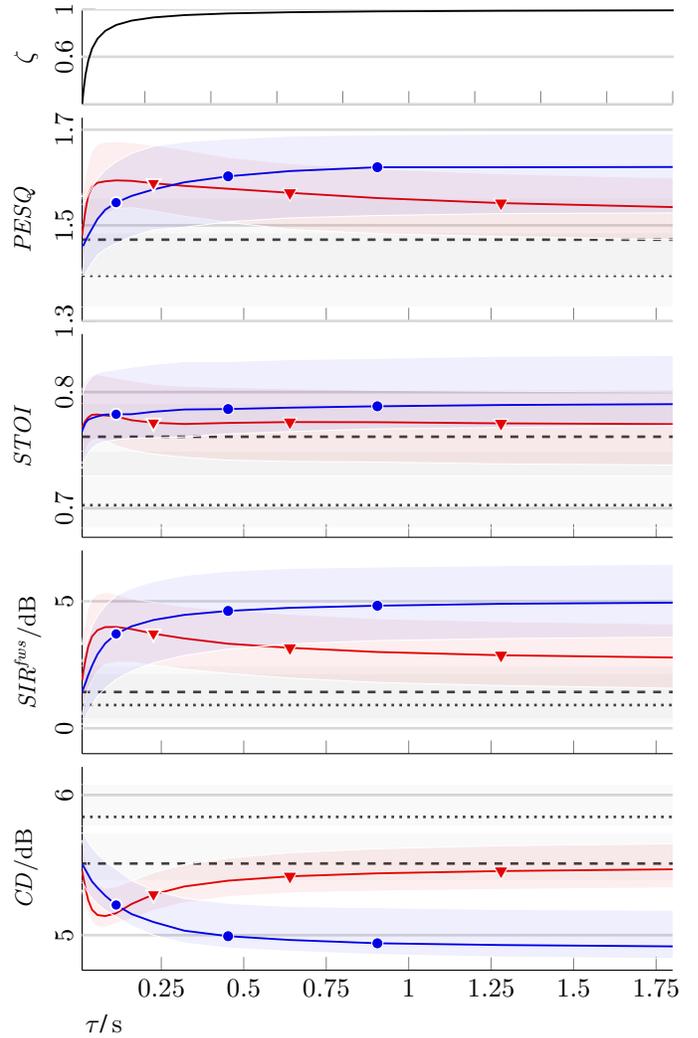} 
\caption{The forgetting factor $\zeta$ [\ref{MWF_tau_1}] and the performance measures $\mathit{PESQ}$, $\mathit{STOI}$, $\mathit{SIR}^{fws}$, and $\mathit{CD}$  versus $\tau$ for the first microphone signal [\ref{MWF_tau_5}],  the MVDR [\ref{MWF_tau_8}], the MWF with smooth PSD estimates [\ref{MWF_tau_13}] and the MWF with instantaneous PSD estimates [\ref{MWF_tau_18}]. 
The graphs denote the median scores over all scenarios, the shaded areas indicate the range from the first to the third quartile.}
\label{fig:sim}
\end{figure}

In this section, we compare the speech enhancement performance of the MWF with smooth PSD estimates $\hat{\varphi}_{s|\textsl{sm}}$, $\hat{\varphi}_{\textsl{d}|\textsl{sm}}$ according to Sec. \ref{sec:smoothEigenvalues} and the MWF with instantaneous PSD estimates  $\hat{\varphi}_{s|\textsl{inst}}$, $\hat{\varphi}_{\textsl{d}|\textsl{inst}}$ according to Sec. \ref{sec:instantPC} as a function of the time constant $\tau$.

In our simulations, we use a linear array of $M=5$ microphones spaced by $8\,\si{cm}$.
In total, $48$ scenarios are generated.
The source is positioned $2\,\si{m}$ away at an angle of $\{0,\,30,\,60\}^\circ$ relative to the broadside direction of the microphone array, where sound propagation is modeled using measured room impulse responses (RIRs) \cite{madivae} of $0.61\,\si{s}$ reverberation time.
In each source position, both male and female speech are used as source signals, where we select $8$ sections of $10\,\si{s}$ from each of the source signal files \cite{bando92}.
Diffuse babble noise \cite{habets2008generating, Audiotec} is added at a signal-to-noise ratio (SNR) of $5\,\si{dB}$, where the SNR is defined as the power ratio of $\x$ and $\v$ in the time domain.
The sampling rate is $f_\textit{s} = 16\,\si{kHz}$.
The STFT processing uses square root Hann windows of $512$ samples with $R=256$ samples overlap.
The presumed available estimates of the RETFs in $\h$ are generated based on the directions of arrival, i.e. the estimate corresponds to the free-field steering vector.
We measure performance in terms of the perceptual evaluation of speech quality $\mathit{PESQ}$ \cite{itu01} with mean opinion scores $\in [1,4.5]$, 
the short-time objective intelligibility $\mathit{STOI}$ \cite{taal11} with scores $\in [0,1]$, 
the frequency-weighted segmental signal-to-interference ratio $\mathit{SIR}^{\mathit{fws}}$ \cite{loizou2007speech} in \si{dB}
and the cepstral distance $\mathit{CD}$ \cite{loizou2007speech} in \si{dB}.
The clean reference signal is generated by convolving the speech source signal with the early part of the RIR to the first microphone.
The computed measures are averaged over all $48$ scenarios. 

Fig. \ref{fig:sim} reports the simulation results.
As to be expected, both versions of the MWF [\ref{MWF_tau_13}, \ref{MWF_tau_18}] outperform the MVDR [\ref{MWF_tau_8}], which in turn shows some improvement over the unprocessed microphone signal [\ref{MWF_tau_5}]. 
The two versions of the MWF however show a different behavior.
The MWF with smooth PSD estimates [\ref{MWF_tau_13}] reaches a fairly sharp performance peak at relatively low values of $\tau$, with decreasing performance for larger values, where the spectral gain in (\ref{eq:MWFgain}) becomes less time-variant.
This behavior is explained by the fact that when computing smooth PSD estimates according to Sec. \ref{sec:smoothEigenvalues}, the time constant $\tau$ trades off the accuracy of the eigenspace basis estimate $\hat{\P}$ on the one hand and the degree of non-stationarity maintained in the PSD estimates $\hat{\varphi}_{s|\textsl{sm}}$ and $\hat{\varphi}_{\textsl{d}|\textsl{sm}}$ on the other hand. 
The MWF with instantaneous PSD estimates [\ref{MWF_tau_18}] in contrast shows a monotonous  performance increase in $\tau$, which facilitates tuning.
This is explained by the fact that when computing instantaneous PSD estimates according to Sec. \ref{sec:instantPC}, the accuracy of the eigenspace basis estimate $\hat{\P}$ still increases with $\tau$, while the instantaneous PSD estimates $\hat{\varphi}_{s|\textsl{inst}}$ and $\hat{\varphi}_{\textsl{d}|\textsl{inst}}$ maintain non-stationarities independently of $\tau$.
At large values of $\tau$, in all measures, the improvement with respect to the MVDR is more than twice as large  for MWF with instantaneous PSD estimates as compared to the MWF with smooth PSD estimates.
Note that in a spatially dynamic scenario with time-varying RETF $\h$, the eigenspace basis $\P$ becomes time-variant, in which case the performance of the MWF with instantaneous PSD estimates might possibly not increase monotonically in $\tau$ anymore, but will presumably show a peak depending on the pace of RETF variations.

\section{Conclusion}
\label{sec:conclusion}

In this paper, as an alternative to smooth PSD estimation based on smooth generalized eigenvalues,
we have proposed an instantaneous PSD estimation approach based on generalized principal components. 
The instantaneous PSD estimates maintain non-stationarities and hence potentially outperform smooth PSD estimates for speech enhancement, as exemplarily shown for the MWF.

\bibliographystyle{IEEEtran}
\bibliography{IEEEabrv,../../../bib/MyResearch}

\end{document}